\documentclass[10pt]{amsart}

\usepackage{hyperref}
\usepackage{graphicx} 

\title[Errico Presutti 1942--2024]{Errico Presutti 1942--2024}

\author[L.\ Bertini]{Lorenzo Bertini}
\address{
Lorenzo Bertini 
\hfill\break\indent
Dipartimento di Matematica, Universit\`a di Roma La Sapienza,
}
\author[J.L.\ Lebowitz]{Joel L.\ Lebowitz}
\address{
Joel L.\ Lebowitz
\hfill\break\indent
Department of Mathematics and Physics, Rutgers University 
}
\author[M.\ Pulvirenti]{Mario Pulvirenti}
\address{
Mario Pulvirenti
\hfill\break\indent
Dipartimento di Matematica, Universit\`a di Roma La Sapienza,
\hfill\break\indent
International Research Center on the Mathematics and Mechanics of
Complex Systems MeMoCS 
\hfill\break\indent
Accademia Nazionale dei Lincei
}

\begin{document}


\maketitle
\thispagestyle{empty}

\begin{quote}
  Errico Presutti was a leading figure in mathematical physics and an
  important contributor to rigorous results in statistical mechanics.
  Due to his strong scientific personality and human qualities, there
  are many who remember Errico Presutti as colleague, mentor, and
  friend.
\end{quote}

\begin{figure}[h]
   \includegraphics[width=0.41\linewidth]{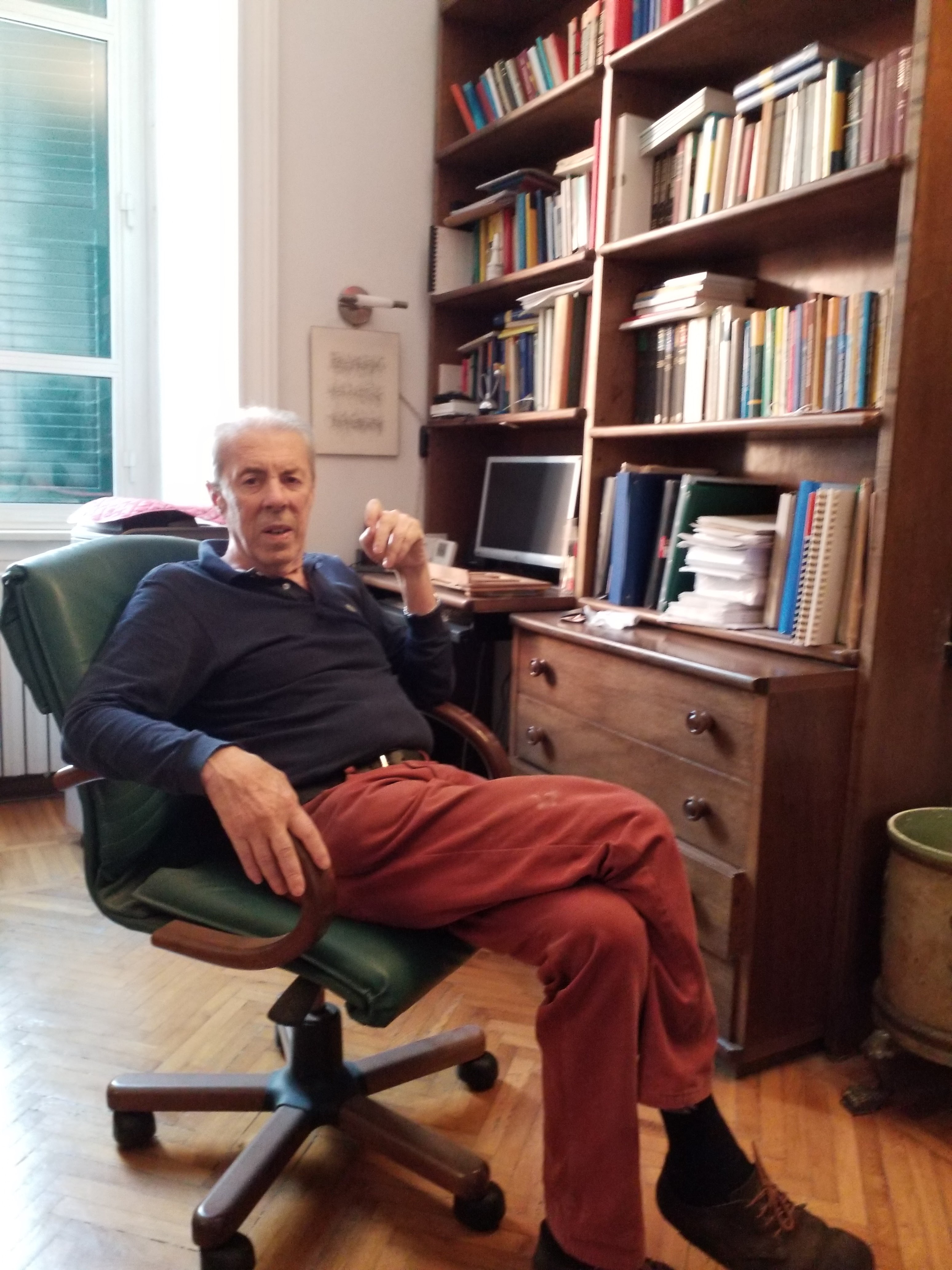}
\end{figure}

\subsection*{Academic carrier}
Errico Presutti graduated in Physics in 1967 at the University of Rome
La Sapienza. At that time there was no PhD in Italy and he became a fellow at the
Physics Institute, Engineering Faculty of Rome La Sapienza.
Between 1971 and 1980 he was Associate Professor, first in the
University of L'Aquila and then in Rome La Sapienza.
In 1980 he was appointed Full Professor in Mathematical Physics. In
this role we was first in L'Aquila, then in Rome La Sapienza, and
finally in Rome Tor Vergata. Retired in 2012, he was Professor Emeritus at Rome Tor Vergata and
active member of GSSI in L'Aquila, playing a key role in starting the
PhD in Mathematics there.

\subsection*{Honors}
\begin{itemize}
\item[1995] Annales Henri Poincar\'e Prize. Best article in
Ann.\ Inst.\ H. Poincar\'e Probab.\ Statist.\ 1995.
\item[2002] Annales Henri Poincar\'e Prize. Best article in Ann.\ Henri
  Poincar\'e 2002.
\item[2005] Prize ``Agostinelli'' in Mathematical Physics. Accademia dei
  Lincei.
\item[2009] Member of  Accademia dei Lincei. 
\item[2014] Prize ``Tullio Levi-Civita'' for the Mathematical and
Mechanical Sciences, M{\&}MoCS.
\end{itemize}

\subsection*{Scientific interests}
The scientific works by E.\ Presutti focused on the
rigorous mathematical approach in statistical mechanics, in the spirit
of Dobrushin, Lanford, Ruelle, Gallavotti, and Sinai.
In his career, he has tackled a large variety of problems including
time evolution of particle systems, Euclidean quantum field theory,
ergodic theory, non-equilibrium statistical mechanics, hydrodynamic
limits, interface dynamics, and phase transitions.
His main emphasis has always been the interplay between the
microscopic and macroscopic description of many body physical systems.
A recurrent theme in E.\ Presutti's activities has been the
introduction in statistical mechanics of novel methods from other
branches of mathematics as operator algebras, functional analysis,
probability, stochastic processes, and geometric measure theory.

\subsection*{Remarkable results}
Referring to the
\href{https://projecteuclid.org/journals/brazilian-journal-of-probability-and-statistics/volume-29/issue-2/Publications-of-Errico-Presutti/bjps/1429105586.full}{list
  of publications by E.\ Presutti}, we
next outline his most relevant results.  

\begin{itemize}
\item[(1970)] In his first publication, E.\ Presutti, in collaboration
  with C.\ Marchioro, discusses the thermodynamic limit of the
  exactly solvable Calogero model.
\item[(1975)] E.\ Presutti shows the equivalence between the
  mechanical notion of the pressure and its thermodynamic
  characterization.
\item [(1975)]
  Together with C.\ Marchioro and A.\ Pellegrinotti,
  he proves the existence of the time evolution for a system of
  infinitely many particles in thermal equilibrium, in any spatial
  dimensions, extending a previous one-dimensional result
  due to O.\ Lanford.
\item[(1978)]
  In collaboration with other Roman colleagues, he shows the
  ultraviolet stability of models in Euclidean quantum field theory by
  establishing a rigorous version of the renormalization group.
\item[(1981)]
  With A.\ Galves, C.\ Kipnis, and C.\ Marchioro, he makes a seminal
  contribution to the field of non-equilibrium statistical mechanics
  by introducing a stochastic microscopic model and solving
  the corresponding stationary non-equilibrium measure, showing in
  particular the validity of Fick's law.
\item[(1985)]
  In collaboration with C.\ Boldrighini, A.\ Pellegrinotti, Ya.G.\
  Sinai, and  M.R.\ Soloviechik, he analyzes the ergodic properties of 
  an infinite system in statistical mechanics.
\item[(1986)]
   Together with A.\ De Masi, H.\ Spohn, and W.D.\ Wick, he
   characterizes the fluctuations of the exclusion processes with
   speed change. 
\item[(1991)]
  In a monograph with A.\ De Masi, he discusses the notion of
  hydrodynamic limits for stochastic systems, providing a rigorous
  definition of local equilibrium. Notice that when the local
  equilibrium is not simply Maxwellian, as for the Boltzmann equation,
  this notion is much more delicate and requires new efforts.
\item[(1993)]
  In a series of articles with A.\ De Masi, E.\ Orlandi, and L.\ Triolo
  he provides a detailed analysis of the Glauber dynamics for Ising
  spin systems with Kac potentials, showing in particular that the
  limiting evolution of interfaces is described by the mean curvature
  flow.
\item[(1995)]
  In collaboration with M.\ Cassandro and R.\ Marra he obtains the
  corrections to the mean field critical temperature for Ising models
  with Kac potentials.
\item[(1995)]
  Together with S.\ Brassesco and A.\ De Masi, he describes the
  interface fluctuations for the stochastic Allen-Cahn equation. 
\item[(1996)]
  Involving colleagues from the De Giorgi's school in calculus of
  variations, he introduced the methods of geometric measure theory in
  statistical mechanics, discussing the Wulff theory for Ising-Kac
  systems.
\item[(1999)]
  Together with J.\ Lebowitz and A.\ Mazel he proves that a system of
  particles in the continuum with finite-range interactions
  exhibits the liquid-vapor phase transition.
\item[(1999)]
  In an influential review with G.\ Giacomin and J.\ Lebowitz he
  discusses the connection between singular non-linear stochastic
  partial differential equations and particle systems.
\item[(2008)] In the monograph `Scaling limits in statistical
  mechanics and microstructures in continuum mechanics',
  he discusses how the continuum description of physical systems
  emerges from the underlying atomistic models.
\item[(2010)]
  With G.\ Gallavotti he introduces microscopic models of thermostats and
  discusses their equivalence in the thermodynamic limit.
\end{itemize}

\subsection*{Influence on younger colleagues}   
The direct influence of Errico Presutti in the mathematical community
is not restricted to his students; he collaborated with many younger 
researchers, who remember him with love and everlasting gratitude.
He treated all his coworkers on equal terms, fully respecting their
autonomy, he essentially never lectured, but he led by example,
continuously suggesting problems and techniques.

\begin{figure}[h]
    \includegraphics[width=0.75\linewidth]{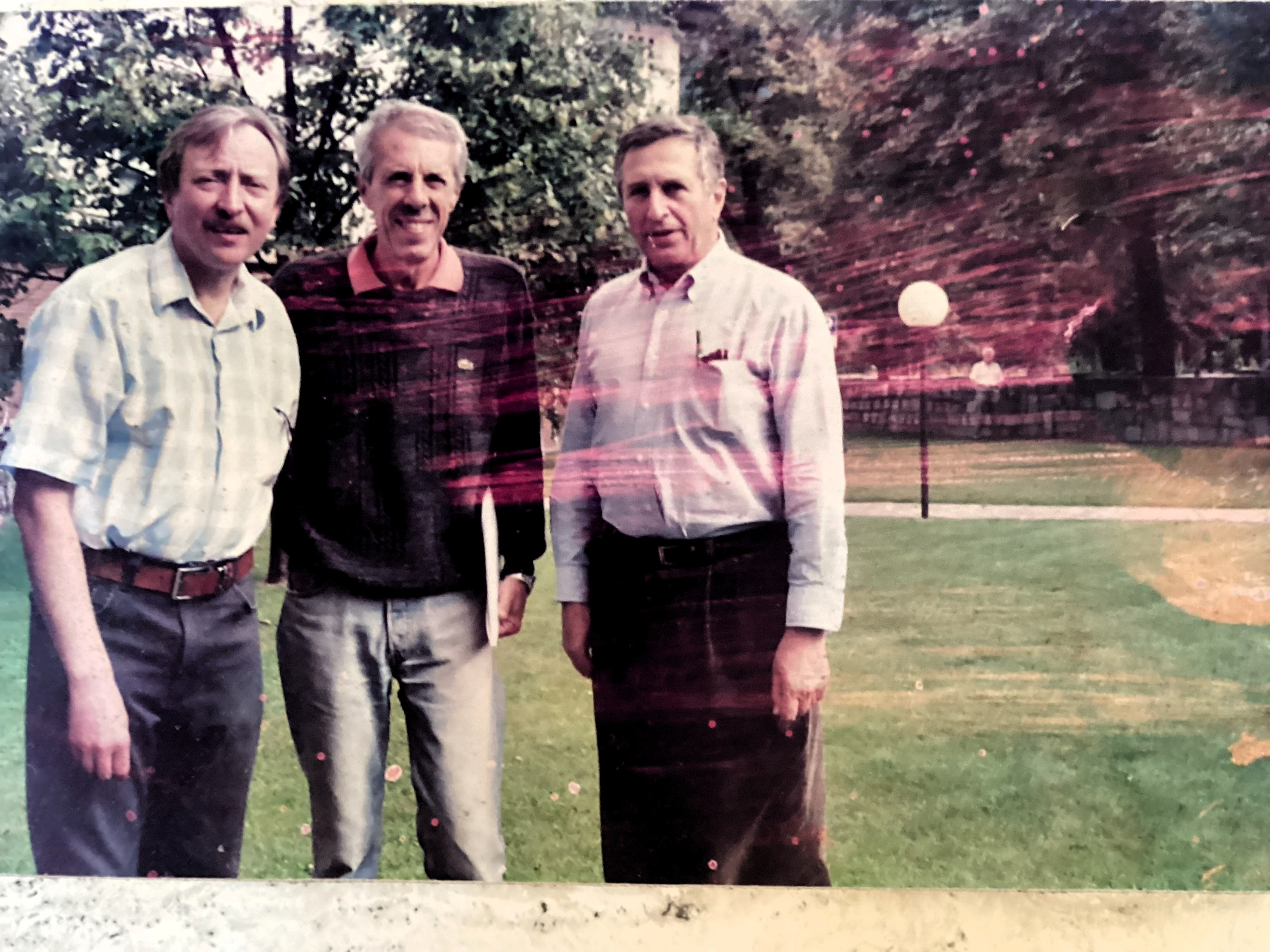}
\end{figure}

\subsection*{My relation with E.\ Presutti (L.\ Bertini)}
In 1991, I started my PhD in Mathematics at the University of Rome
Tor Vergata, and I first met Errico. Within a couple of weeks, he had
me involved in a research project, the analysis of the critical
fluctuations of the Glauber dynamics for a one-dimensional Ising model
with Kac potentials.
This project lasted for about one year, and the number of topics that
I learned from Errico is the overwhelming majority of what I know
today: martingales, particle systems, stochastic equations,
statistical mechanics, and partial differential equations. Yet, his
greater lesson was another: he transferred to me his passion for
science and the love of concreteness in mathematics. If almost forty
years later I am still able to do some original research, it is for
the little flame he installed in my heart.

We then agreed my PhD thesis should be about the exclusion process
and the stochastic Burgers equation. I expressed, however, the wish to
spend a year in the US that he fully supported. After a while, I got a
clearer idea of what I would have liked to do and sent him a
message. In his answer, he explained in detail the steps to achieve my
goals; the final outcome was a joint paper with G.\
Giacomin, ``Stochastic Burgers and KPZ equations from particle
systems'', that I regard as my most original contribution.

I had other occasions to collaborate with Errico, but his influence on
my activities goes far beyond that. I discussed with him essentially
every problem that I tackled and he provided crucial insight, always
accompanied by his secret smile hiding behind his mustache.

\subsection*{My relation with E.\ Presutti (J.L.\ Lebowitz)}   
I first met Errico in 1975 when he came to visit me at Yeshiva
University in New York City. He had just driven across the US,
starting at Stanford, where we was visiting Donald Ornstein. I believe
that both visits were arranged by Giovanni Gallavotti. I also recollect
that he stopped during his cross country drive in Las Vegas for a
concert by Frank Sinatra. We hit it off during that visit and became
lifelong friends and collaborators.

We published a dozen of papers or so, mostly also with other
collaborators, between 1976 and 2023. During this period we also spent
literally thousands of hours discussing science. This was during my
visits to Rome when I stayed in his home, during his extended visits
to Rutgers and during our joint stay at the IHES in Bures-sur-Yvette.
In fact Errico should have been a co-author on many other papers I
published and that of many others, if it was not for his insistence of
remaining in the background.

The work I did with Errico involved both equilibrium and
nonequilibrium systems. His real love, as he describes it in the
preface to his wonderful book, ``Scaling
Limits in Statistical Mechanics and Microstructures in Continuum Mechanics'', is
``The way a continuum description emerges from atomistic models is an
intriguing and fascinating subject which is behind most of my
scientific life.''

In this book Errico gives a detailed exposition of some work we did
jointly with Alex Mazel proving for the first time the existence of a
liquid vapor transition in a continuum systems with finite range
interactions and no symmetries. To do this required the introduction of
a four body repulsive interaction. The proof of such a transition for
systems with only pair interactions is still missing. It is a serious
gap in the rigorous mathematical theory of phase transitions, which I
hope will be closed soon. To quote from the article in PRL: ``An outstanding problem in equilibrium
statistical mechanics is to derive rigorously the existence of liquid-vapor phase transition
in particles interacting with any kind of reasonable potential, say Lennard-Jones or hard
core plus attractive square well ... In this letter we report the first proof of a liquid-vapor
transition in one-component continuum systems with finite range interactions and no
symmetries. The basic idea of our approach is to study perturbations not of the ground
state but of the mean field behavior (mfb): i.e., we shall consider situations where the
interactions are 
parametrized by their range $1/\gamma$ and perturb around $\gamma=0$
which gives the mean field behavior.''

\subsection*{My relation with E.\ Presutti (M.\ Pulvirenti)}   

I know Errico Presutti since the beginning of our scientific
careers. We had a common interest in statistical mechanics in the
group formed in Roma by G.\ Gallavotti.

In the course of the years we have not written many papers together
but we spent a lot of time in scientific discussions, plans and
hopes. Not only, we also did common activities in real life, like to
play at the soccer game or bridge. Incidentally Errico was a very good
player in both.

I learnt a lot from Errico as regards the rigorous analysis of the
scaling limits in non-equilibrium statistical mechanics and the
derivation of effective macroscopic equations. In particular Errico
explained me

1) Superstability estimates. I had difficulties in understanding it
well. Errico showed me that, assuming preliminary that the potential
is short range, then the proof was simpler and conceptually clear.

2) $v$-functions. In proving hydrodynamic limits often one has to
compare two family of marginals, say $f_j$ and $g_j$. Suppose for the
moment that they factorize. Then the difference $f_j - g_j$ is much
larger than the product of the differences.  Say $\varepsilon$ against
$\varepsilon^j$. Expanding such a product one obtain a definition
($v$-functions) which can be used also if $g_j$ do not factorizes. This
allows more accurate estimates in problems concerning macroscopic
limits. Errico explained me this and I used this notion (which is a
sort of reduced cumulants) in the study of kinetic limits.

3) Doeblin condition. Errico explained me it in a very simple
way. This was very useful.

I am deeply impressed by the Errico generosity in following a lot of
young people, giving ideas and help without appearing himself, but
promoting their scientific autonomy.

$~$\bigskip$~$

\end{document}